\documentclass[pra,aps,twocolumn,superscriptaddress,showpacs,amsmath,amssymb,floatfix]{revtex4}
\usepackage{comment}
\usepackage{graphicx}
\usepackage{epstopdf}

\begin{document}

\title{Effects of interaction on the diffusion of atomic matter waves \\ in one-dimensional 
quasi-periodic potentials}

\author{M. Larcher}
\affiliation{CNR INFM-BEC and Dipartimento di Fisica, Universit\`a di Trento, 38050 Povo, Italy}
\author{F. Dalfovo}
\affiliation{CNR INFM-BEC and Dipartimento di Fisica, Universit\`a di Trento, 38050 Povo, Italy}
\author{M. Modugno}
\affiliation{LENS and Dipartimento di Fisica, Universit\`a di Firenze, Via N. Carrara 1,50019 Sesto Fiorentino, Italy}

\date{September 9, 2009}

\begin{abstract}
We study the behaviour of an ultracold atomic gas of bosons in a bichromatic lattice, where the weaker lattice is used as a source of disorder. We numerically solve a discretized mean-field equation, which generalizes the one-dimensional Aubry-Andr\`e model for particles in a quasi-periodic potential by including the interaction between atoms. We compare the results for commensurate and incommensurate lattices. We investigate the role of the initial shape of the wavepacket as well as the interplay between two competing effects of the interaction, namely self-trapping and delocalization.  Our calculations show that, if the condensate initially occupies a single lattice site, the dynamics of the interacting gas is dominated by self-trapping in a wide range of parameters, even for weak interaction. Conversely, if the diffusion starts from a Gaussian wavepacket, self-trapping is significantly suppressed and the destruction of localization by interaction is more easily observable. 
\end{abstract}

\maketitle

\section{Introduction}

Localization induced by disorder has been recently observed in ultracold bosonic gases in purely random potentials \cite{Aspect} and in bichromatic optical lattices \cite{Roati}. In the experiment of Ref.~\cite{Aspect}, a Bose-Einstein condensate of interacting atoms is released from a trap into a speckle potential and the localization effects appear in the low density tails of the spatial distribution of atoms, where the interaction is negligible. In \cite{Roati}, the interaction between atoms is suppressed from the very beginning by tuning the $s$-wave scattering length to zero by means of a Feshbach resonance. In both cases, the observations have been interpreted in terms of Anderson localization \cite{Anderson}. 

An interesting feature of one-dimensional (1D) bichromatic lattices is that they allow one to implement the Aubry-Andr\`e model \cite{Aubry}, also known as Harper model \cite{Harper}, for particles in a quasi-periodic potential. In the case of noninteracting particles, this model predicts a sharp transition from diffusion to localization for a given value of the disorder strength.  In ultracold gases, thanks to the availability of Feshbach resonances, the interaction between atoms can be changed almost at will, thus allowing for the investigation of the role played by interaction in the transition from diffusion to localization \cite{Florence-preprint}. The effects of interaction have been recently the subject of several theoretical investigations in the case of localization in purely random potentials \cite{Kopidakis, Pikovsky,  Garcia, Flach, Skokos, paul2009} and quasi-periodic potentials \cite{deng2009,adhi2009,ng2009}, but some results are still controversial.  It is worth mentioning that the Aubry-Andr\`e model has been recently implemented also in experiments with diffusion of light in photonic lattices \cite{lahini2008direct}.

The purpose of this paper is to understand the interplay between diffusion and localization of an interacting Bose-Einstein condensate in a 1D bichromatic lattice. One of the main problems arises from the occurrence of two competing effects of the interaction: on the one hand it favours localization through the self-trapping mechanism \cite{trombettoni2001}, on the other hand it is expected to destroy the (Anderson) localization induced by disorder. In order to shed light on these effects and make a bridge between theoretical results and feasible experiments, we study the following issues: i)~what happens when the two lattices, which are superimposed to create the bichromatic lattice, have commensurate wavelengths  (periodic system) instead of incommensurate (quasi-periodic system, as in the Aubry-Andr\`e model); ii) what is the role of the initial shape of the condensate; iii) what is the role of self-trapping processes. All these issues have subtle implications in the experimental observability of the crossover between diffusion and localization in an interacting gas. Our calculations show that, if the condensate initially occupies a single lattice site, the dynamics of the gas in the lattice is dominated by self-trapping in a wide range of parameters, even for weak interaction. Conversely, if the diffusion starts from a condensate with Gaussian shape, extended over several lattice sites, self-trapping is significantly suppressed and the destruction of localization by interaction is more easily observable. 

\section{The model}\label{sec:model}

One-dimensional bichromatic lattices are realized in experiments with Bose-Einstein condensates by superimposing two optical lattices of different wavelengths, producing an external potential acting on the atoms in this form: 
\begin{equation}
V(x)=s_1 E_{R_1}\cos^2(k_1x)+s_2 E_{R_2}\cos^2(k_2x) \, , 
\end{equation}
where $E_{R_j}=\hbar^2 k_j^2/(2m)$ is the recoil energy and $s_j$ is the dimensionless lattice strength. One of the two lattices is used as the main periodic potential (primary lattice) fixing the Bloch band structure of the single-particle states without disorder. It is usually strong enough ($s_1\gg 1$) to apply the tight-binding approximation, i.e., the atoms occupy the sites of the primary lattice and can tunnel from one site to the other with a given tunneling rate $J$ \cite{modugno2009,Biddle}. The second lattice is weaker ($s_2\ll s_1$) and introduces a ``deterministic" disorder, or quasi-disorder \cite{Roati, lye2007, fallani2007}. For a noninteracting gas in such a potential, the evolution of the system is described by the Aubry-Andr\`e model \cite{Aubry}, which is obtained from the Schr\"odinger equation by expanding the single-particle wavefunction $\psi(x)$ over a set of Wannier states, maximally localized at the minima of the primary lattice in the lowest Bloch band, $|\psi\rangle=\sum_j \psi_j|w_j\rangle$ \cite{boers2007, modugno2009}. In the presence of interactions between the atoms, one can instead start from the Gross-Pitaevskii (GP) equation \cite{Pita, Gross} and use the same procedure in order to get a generalized Aubry-Andr\`e model which includes an additional nonlinear term that represents the mean-field interaction. The Hamiltonian is
\begin{equation}
\label{eq:Hamiltonian}
H=\sum_j-(\psi_{j+1}\psi_j^*+\psi_{j+1}^*\psi_j)+V_j |\psi_j|^2+\frac{1}{2}\beta|\psi_j|^4,
\end{equation}
with 
\begin{equation}\label{eq:potential}
V_j=\lambda \cos(2\pi\alpha j+\theta),
\end{equation}
where $j$ is the primary lattice site index, $\alpha=k_2/k_1$ is the ratio between the wavenumbers of the two lattices, $\theta$ is an arbitrary phase, $\psi_j$ is a complex variable whose modulus square gives the probability of finding a particle at the lattice site $j$. In deriving the Hamiltonian (\ref{eq:Hamiltonian}) from the GP equation for a realistic three-dimensional condensate, we assume the condensate to be axially symmetric and elongated in the direction of the lattice, such that the transverse confinement is sufficiently strong to freeze out the radial dynamics and the longitudinal confinement sufficiently weak to be neglected. We have also chosen $E_{R_1}J=1$. The dimensionless parameters $\lambda$ and $\beta$ represent the strength of the disorder and of the mean-field interaction, respectively, and are the key parameters of the present work. 
 
The equations of motion are generated by $i{\partial \psi_j}/{\partial t}={\partial H}/{\partial \psi_j^*}$, yielding
\begin{equation}\label{eq:NLSE}
i\frac{\partial\psi_j}{\partial t}=-\psi_{j+1}-\psi_{j-1}+V_j \psi_j+\beta|\psi_j|^2\psi_j
\end{equation}
that can be considered as the GP equation on a discretized lattice. Similar versions of a discretized GP equation have been already used to investigate the dynamics of condensates in different situations (see for instance Ref.~\cite{trombettoni2001}). In the above equation, the time $t$ is expressed in dimensionless units. The actual time in seconds can be obtained by multiplying $t$ by $\hbar/(JE_{R_1})$.

We study the localization properties of the system by considering the problem of quantum diffusion of an initially localized wavepacket. We solve Eq.~(\ref{eq:NLSE}) by using a standard fourth order Runge-Kutta algorithm for the numerical integration. The accuracy of the integration is checked by monitoring the conservation of the norm of the wavepacket and of the energy of the system. We investigate the evolution starting from two different classes of initial conditions, namely a $\delta$-function localized in a single lattice site,
\begin{equation}
\psi_j(0)=\delta_{j,0} \, ,
\end{equation}
and a Gaussian wavepacket of width $\sigma$,
\begin{equation}
\psi_j(0)=Ne^{-\frac{j^2}{2\sigma^2}} \, ,
\end{equation}
where $N$ is a normalization factor that has to be determined according to $\sum_j |\psi_j|^2=1$. Owing to arbitrariness of the phase $\theta$, here we have chosen, without any loss of generality,  to fix the initial localization center at $j=0$.

As a measure of the localization we consider two quantities: the width of the wavepacket measured as the square root of the second moment of the spatial distribution $|\psi_j(t)|^2$, 
\begin{equation}
w(t)=\sqrt{m_2(t)}= \{ \sum_j(j-\langle j\rangle)^2|\psi_j(t)|^2 \}^{1/2} \, ,
\label{eq:width}
\end{equation} 
and the participation number 
\begin{equation}
P(t)=\frac{1}{\sum_j|\psi_j(t)|^4} \, , 
\end{equation}
which measures the number of significantly occupied lattice sites \cite{Ingold}. The quantity $\langle j\rangle$ represents the average over the spatial distribution $|\psi_j(t)|^2$, defined as $\langle j\rangle=\sum j|\psi_j|^2$.

\begin{figure}[b!]
\begin{center}
\includegraphics[width=0.95\columnwidth]{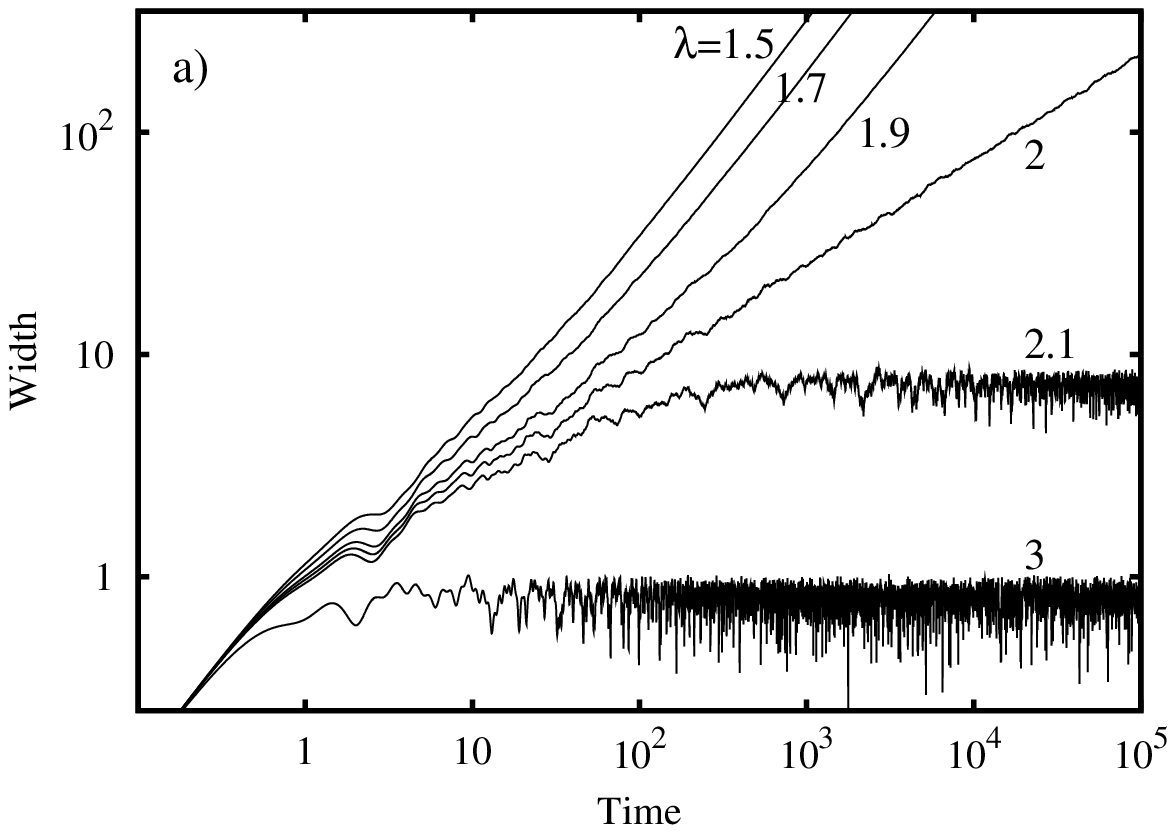}
\includegraphics[width=0.95\columnwidth]{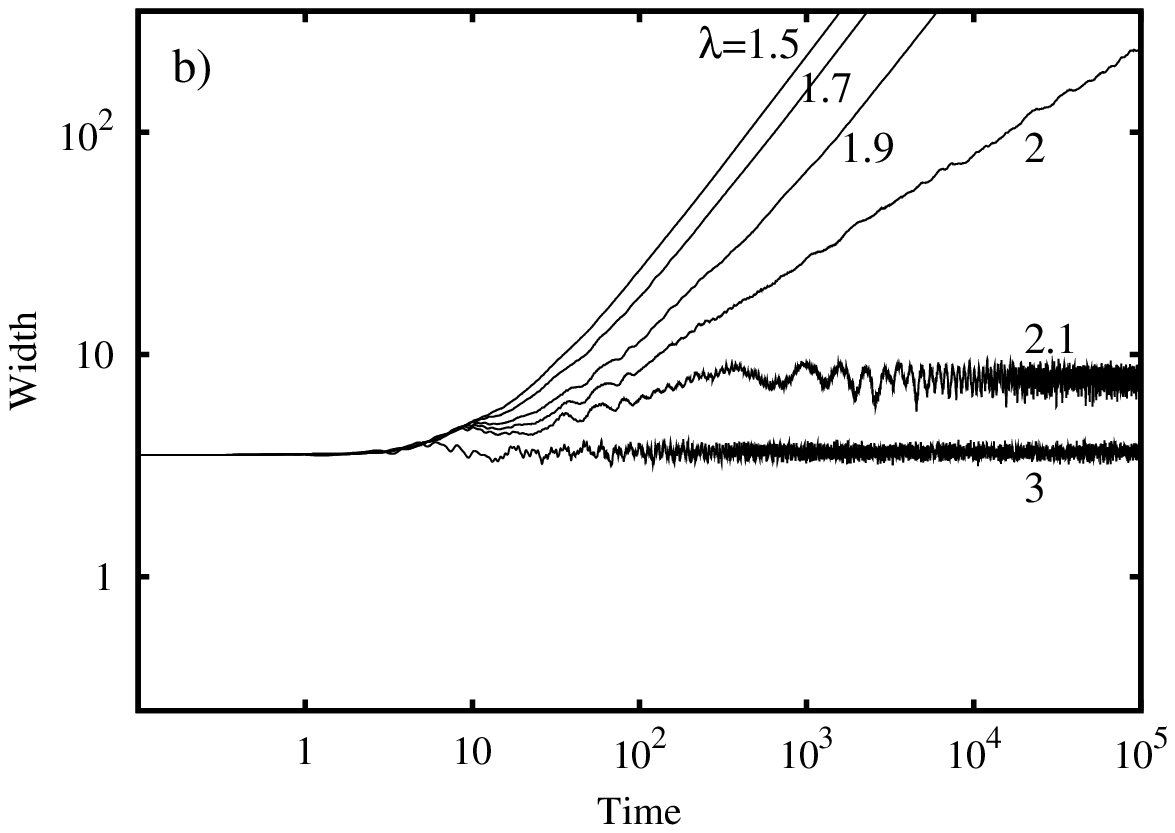}
\end{center}
\caption{Diffusion of a noninteracting gas ($\beta=0$) in the Aubry-Andr\`e model with $\alpha=(\sqrt{5} - 1)/2$. The time evolution of the width of the wavepacket $w(t)$ is shown for different values of the disorder strength, $\lambda=1.5, 1.7, 1.9, 2, 2.1, 3$. In the top panel, the starting wavepacket is a $\delta$-function localized in a single site. In the bottom panel we use an initial Gaussian wavepacket of width $\sigma= 5$. In both cases, one clearly observes the transition from extended to localized states that occurs at $\lambda=2$. Here and in the following figures, time and width are expressed in the dimensionless units defined in section \ref{sec:model}.  }
\label{fig:non-interacting}
\end{figure}

\section{Noninteracting particles}

When $\beta=0$ (noninteracting gas) and $\alpha$ is irrational (quasi-periodic lattice),  Hamiltonian (\ref{eq:Hamiltonian}) coincides with the well-known Aubry-Andr\`e model that has been widely studied in the literature (see \cite{modugno2009} and references therein). Aubry and Andr\`e \cite{Aubry} analytically proved that the system undergoes a transition from extended to localized states at $\lambda=2$. All states are extended for $\lambda<2$, critical for $\lambda=2$ and exponentially localized for $\lambda>2$, with the same localization length $1/{\ln(\lambda/2)}$. Aubry and Andr\`e \cite{Aubry} identified this exponential localization as the Anderson localization in a quasi-periodic potential, analog to Anderson localization in a purely random potential \cite{Albert}. 

From the numerical viewpoint, the static spectrum of the Aubry-Andr\`e Hamiltonian can be calculated by solving the following eigenvalue problem
\begin{equation}\label{eq:aubry_andre}
-\psi_{j+1}-\psi_{j-1}+\lambda \cos(2\pi\alpha j+\theta) \psi_j=E\psi_j
\end{equation} 
and considering a sequence of rational numbers $\alpha_n$, that converges to an irrational $\alpha$ as $n\rightarrow\infty$ (see for instance \cite{Tang, Hiramoto_2}). The sequence of approximants $\alpha_n$ can be found by successive truncations of the continued-fraction expansion of $\alpha$. A standard choice consists of choosing the inverse golden mean $\alpha=(\sqrt{5} - 1)/2$ \cite{Ingold}. In this case the approximants are obtained by writing $\alpha_n=p_n/q_n$, where $p_n$ and $q_n=p_{n+1}$ are two consecutive terms of the Fibonacci sequence ($p_1=p_2=1$, $p_{n}=p_{n-1}+p_{n-2}$ for $n>2$). The incommensurate case can thus be considered as the limit of a sequence of commensurate Hamiltonians, whose eigenvalues $E^{k,m}$ and eigenfunctions $\phi^{k,m}_j$ can be labelled by the quasi-momentum $k$ and the band index $m$, since the spatial periodicity of the system, with period $q_n$, permits to use the Bloch wave decomposition. One finds that, for sufficiently large $n$ and for  $\lambda>2$, the eigenfunctions are indeed characterized by periodic replica of exponentially localized functions within each period of the potential, that in the limit $n\rightarrow\infty$ tend to a single localized function \cite{modugno2009}.

The localization transition at $\lambda=2$ can be observed also in the dynamics (quantum diffusion), by looking for example at the width of the wavepacket as a function of the time \cite{Hiramoto}. In particular, in the quasi-periodic lattice, the asymptotic spreading of the wavepacket width $w(t)$ can be parametrized as $w(t)\sim t^{\gamma}$, and one finds three different regimes:
\begin{itemize}
\item[(i)]{$\lambda<2$:} ballistic regime, $\gamma=1$
\item[(ii)]{$\lambda=2$:} sub-diffusive regime, $\gamma\sim0.5$
\item[(iii)]{$\lambda>2$:} localized regime, $\gamma=0$  \, .
\end{itemize}
Our results for $\alpha=(\sqrt{5} - 1)/2$ are shown in Fig.~\ref{fig:non-interacting}. In the case of an initial $\delta$-function wavepacket (top panel), we find perfect agreement with previous calculations \cite{Hiramoto}. The lower panel shows our results for the case of an initial Gaussian wavepacket. By comparing the two cases, one can see that the asymptotic behaviour is not affected by the choice of the initial shape of the wavepacket. 

\begin{figure}[t!]
\begin{center}
\includegraphics[height=0.95\columnwidth,angle=-90]{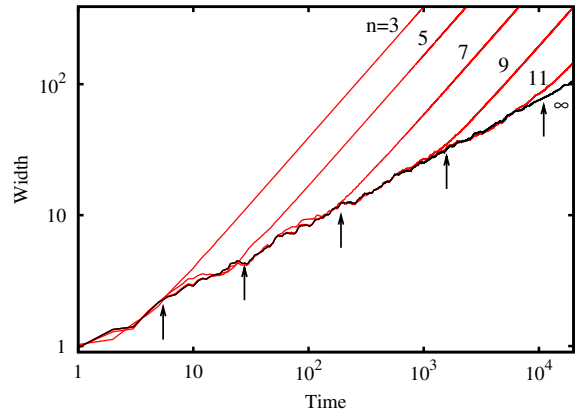}
\end{center}
\caption{(color online) Time evolution of the width of the wavepacket $w(t)$ of noninteracting particles, starting from a single-site $\delta$-function, for $\lambda=2$ and for different orders, $n$, of the approximants in the Fibonacci sequence. The black arrows represent the values of $t$ at which we observe the transition from the behaviour predicted for a quasi-periodic potential (incommensurate lattice)  to the diffusion expected in a periodic potential. } 
\label{fig:time}
\end{figure}

It is worth stressing that a truly quasi-periodic potential can not be realized in any realistic experiment with ultracold atoms,
since the system is finite and the ratio of the frequencies of the two laser beams is a rational number.
It is thus important to clarify to which extent the predictions of the Aubry-Andr\`e model are relevant for the description of current experiments
in bichromatic lattices.  To this purpose we compare the results of the quasi-periodic potential with those obtained by using approximants of order $n$. For the value $\alpha=(\sqrt{5} - 1)/2$, this consists of selecting the term of order $n$ in the Fibonacci sequence. For any finite value of $n$ the system is periodic, with wavelength $q_n$, and the diffusion of an initially localized wavepacket is expected to be ballistic ($w(t)\sim t$). However, in the limit $n\to \infty$ one must recover the results of the Aubry-Andr\`e model, with a critical behaviour for $\lambda=2$ and localized states for $\lambda >2$. The approach to this limit in nontrivial and  involves the characteristic time and length scales of the system. 

In Fig.~\ref{fig:time} we first show our results for the diffusion of a $\delta$-like wavepacket in a lattice with the critical value $\lambda=2$. For any finite $n$ the wavepacket exhibits a sub-diffusive spreading ($w(t)\sim t^\gamma$ with $\gamma\approx0.5$), as in the incommensurate case, within an initial time interval. Then, at time $\tau$, the width starts growing as in a ballistic diffusion in a periodic lattice. The transition between the two regimes turns out to occur when the width of the wavepacket becomes of the same order of the spatial periodicity of the lattice. The transition time, $\tau$, indicated by the arrows in Fig.~\ref{fig:time}, increases with the order $n$ of the approximants and the corresponding width, $w(\tau)$ exhibits a linear dependence on the periodicity of the system, $q_n$ \cite{note-time}. 

\begin{figure}[t!]
\begin{center}
\includegraphics[width=0.95\columnwidth,angle=0]{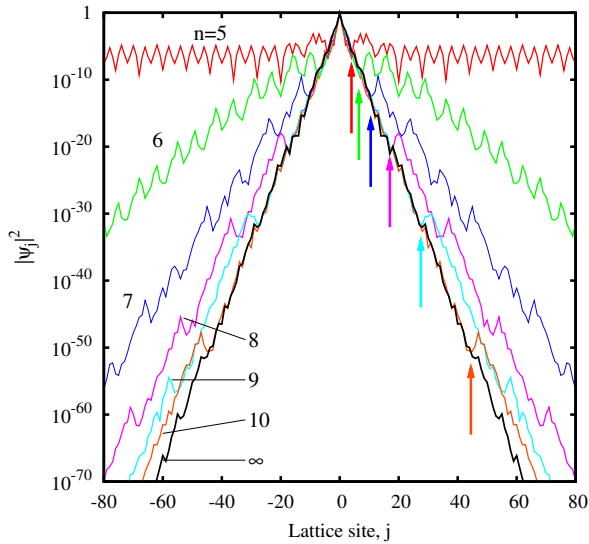}
\end{center}
\caption{(color online) Modulus square of the wavefunction $|\psi_j|^2$ for different values of $n$, plotted at a fixed evolution time $t=1000$, for $\lambda=7$ and $\beta=0$. The initial wavepacket at $t=0$ is a $\delta$-function localized at $j=0$. The vertical arrows are drawn at the positions $q_n/2$.}
\label{fig:wf_comm_incomm_7}
\end{figure}

\begin{figure}[b!]
\begin{center}
\includegraphics[height=0.95\columnwidth,angle=-90]{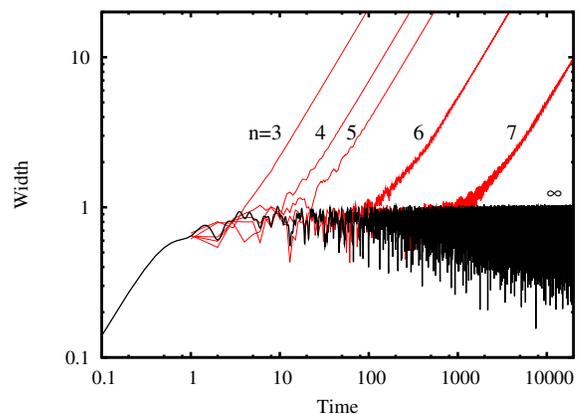}
\end{center}
\caption{(color online) Time evolution of the width of the wavepacket $w(t)$ of noninteracting particles, starting from a single-site $\delta$-function, for $\lambda=3$ and for different values of $n$.}
\label{fig:comm_incomm_3}
\end{figure}

The role of the spatial periodicity is even more evident if one plots the density distribution in the regime of localization, as shown in Fig.~\ref{fig:wf_comm_incomm_7} for $\lambda=7$ and $t=1000$. In this figure the arrows are drawn at the positions $q_n/2$. As one can see, the deviations from the density distribution of the incommensurate case ($n \to \infty$) are caused by the spreading of the lateral components of the distribution, i.e., those at a distance of the order of, or larger than $q_n/2$. The asymptotic behaviour ($t \to \infty$) is always ballistic. However, for a finite $t$ and for $\lambda>2$ the central part of the density distribution (within a width of order $q_n$) exhibits an exponential localization, independent of $n$, and is almost indistinguishable from the one predicted for the incommensurate lattice. The spreading of the low density tails affects the behaviour of the width defined in Eq.~(\ref{eq:width}). An example is shown in Fig.~\ref{fig:comm_incomm_3}. For short times the contribution of the expanding tails is negligible, while for later times the width increases as in a ballistic diffusion. It is worth stressing, however, that these effects of the low density tails are expected to be hardly detectable in actual experiments, due to the finite resolution in the measurement of the density distribution. 

Given the typical timescale \cite{note-expt} and optical resolution of the experiments with ultracold gases in optical lattices, our analysis confirms that the transition from diffusion to localization observed in  Ref.~\cite{Roati} can correctly be interpreted in terms of the predictions of the Aubry-Andr\`e model.

\section{Interacting particles}

In this section we consider the interacting case $(\beta\neq 0)$. We mainly focus on two effects of the interaction, namely the self-trapping phenomenon and the delocalization induced by the interaction in the regime $\lambda>2$. As already mentioned in the introduction, these two effects are competing and must be carefully analysed in order to correctly interpret the expansion of a wavepacket. 

\begin{figure}[t!]
\begin{center}
\includegraphics[height =0.95\columnwidth,angle=-90]{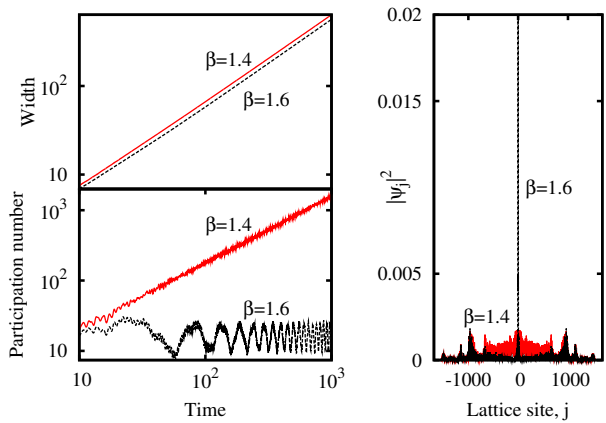}
\end{center}
\caption{(color online) Width $w(t)$, participation number $P(t)$, and density distribution $|\psi_j(t=1000)|^2$ for two values of the interaction strength $\beta$, below ($\beta=1.4$, red lines) and above ($\beta=1.6$, black lines) the transition from diffusion to self-trapping. Here the initial state is a single-site $\delta$-function with $\theta=0$ and $\lambda=0.8$.}
\label{fig:self_trapping}
\end{figure}

\begin{figure}[b!]
\begin{center}
\includegraphics[height=1\columnwidth,angle=-90]{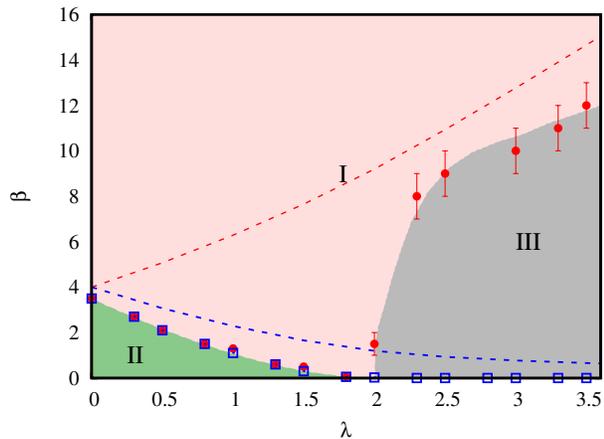}
\end{center}
\caption{(color online) Critical value of the interaction strength for the transition to self-trapping, $\beta_c$, as a function of the disorder strength, $\lambda$, for diffusion from a single lattice site and for $\theta=0$ (blue squares) and $\theta=\pi$ (red circles). The red and blue dashed lines are the corresponding upper bounds for $\beta_c$ obtained by calculating the bandwidth of the single-particle spectrum and using energy conservation arguments. The diagram is schematically divided in three regions I, II and III. All states in I are self-trapped; in II, one finds diffusion, with soliton-like structures and discrete breathers; in III, the transition from diffusive states to self-trapping strongly depends on the value of the phase $\theta$ (i.e., the position of the initial wavepacket). The semi-axis $\lambda>2$ and $\beta=0$ corresponds to the regime of disorder induced localization for noninteracting particles.}
\label{fig:self_trapping_2}
\end{figure}

\subsection{Self-trapping}
\label{sec:self_trapping}

Self-trapping is a localization phenomenon, different from Anderson localization, that occurs when the interaction is stronger than a critical value $\beta_{c}$, even for a purely periodic system without disorder \cite{trombettoni2001, anker2005, Johansson, rosen2008} and double-well potentials \cite{st_double_well_th, st_double_well_exp}. An intuitive understanding of the origin of the self-trapping in a lattice is based on energy conservation arguments \cite{Kopidakis}. Let us consider separately the contribution to the energy that comes from the kinetic and potential terms together and the contribution that comes from the interacting term, $H(t)=H_{0}(t)+H_{int}(t)$. If the gas is subject to a periodic potential in the tight-binding approximation and its dynamics is restricted to the lowest Bloch band, as supposed in deriving Eq.~(\ref{eq:Hamiltonian}), the term $H_0$ in the Hamiltonian is upper bounded. Let us call this upper bound $E_0^{max}$. Whenever the initial energy of the interacting system is larger than this upper bound, $H(t=0)>E_0^{max}$, one can prove that the system cannot reach a situation where $H_{int}(t)=0$, at any $t>0$, without violating the energy conservation. This means that, under these conditions, part of the interaction energy must be trapped in the system in the form of a localized peak that does not spread. In other words, whenever $H(t=0)>E_0^{max}$ an initially localized wavepacket cannot spread to zero in the whole space. This argument, in general, does not provide a precise quantitative estimate of the critical value $\beta_c$, but it gives a reasonable upper bound.

Self-trapping of particles in a 1D quasi-periodic potential for $\lambda<2$ has been already discussed in Ref.~\cite{Johansson} (for the case of purely random potential see \cite{Flach, Skokos, Kopidakis}). Here we provide a more systematic calculation of $\beta_c$ and we compare the diffusion from a single-site to the one from a Gaussian wavepacket.  

A signature of the presence of self-trapping is a saturation of the participation number $P(t)$ that, for $\beta>\beta_c$, reaches an asymptotic finite value, due to the trapping mechanism occurring at the center of the wavepacket, while the width $w(t)$ keeps increasing owing to the expanding tails \cite{self_trapping}.  An example of self-trapping transition is shown in Fig.~\ref{fig:self_trapping}, where we show the results obtained by solving Eq.~(\ref{eq:NLSE}) for diffusion from a single-site in a quasi-periodic potential with $\alpha=(\sqrt{5} - 1)/2$. In the figure one can see the typical change of behaviour that occurs when $\beta$ crosses the critical value $\beta_c$. The same figure shows also the difference in the density distributions at $t=1000$: the central peak is well visible in the self-trapped state and it is absent in the diffusive state, while the lateral, low density tails are similar. 

By systematically looking at the numerical results for $w(t)$, $P(t)$, $|\psi_j(t)|^2$ in the $\beta$ {\it vs.} $\lambda$ plane, we obtain the diagram shown in Fig.~\ref{fig:self_trapping_2}. The values of $\beta_c$ are represented by red circles and blue squares for $\theta=\pi$ and $0$, respectively. In region I, above the red circles, all points correspond to self-trapped states. For $\lambda<2$ we find that the value of $\beta_c$ is practically independent of the phase $\theta$ and decreases as $\lambda$ is increased. In region II, we observe diffusion, often accompanied by solitonic structures and discrete breathers eventually spreading. Similar structures in the numerical solutions of Eq.~(\ref{eq:NLSE}), for diffusion from a single-site and for $\lambda=0$, have been already found in Ref.~\cite{Johansson}. For $\lambda>2$ we find that $\beta_c$ is strongly $\theta$-dependent. In the figure we show the results for the two limiting values $\theta=0$ and $\theta=\pi$; in particular, in region III, we find that all states are self-trapped for $\theta=0$ while they are diffusive for $\theta=\pi$. The semi-axis $\lambda>2$ and $\beta=0$ corresponds to the regime of disorder induced localization for noninteracting particles.

The phase dependence of $\beta_c$ for $\lambda>2$ can be qualitatively explained by the energy conservation arguments already mentioned above. In particular, we numerically calculate the maximum energy, $E_0^{max}$, in the lowest Bloch band of the noninteracting single-particle spectrum and we compare this value to the initial energy of the interacting system, which is given by $H(t=0)=\lambda\cos(\theta)+\beta/2$. The upper bound for the transition to self-trapping is then given by the condition $H(t=0)=E_0^{max}$, which implies 
$$
\beta=2 (E_0^{max}\pm \lambda)
$$
where the plus and the minus signs holds for $\theta=\pi$ and $\theta=0$, respectively. These two upper bounds are represented by the blue and red dashed lines in Fig.~\ref{fig:self_trapping_2}. 

Fig.~\ref{fig:self_trapping_2} shows that, in the case of diffusion from a single-site, the self-trapping mechanism plays a rather important and nontrivial role, leaving almost no space to the observability of the interplay between disorder and interaction. The region were this interplay might be observed, namely for $\lambda>2$ and small $\beta$, where one expects to see the destruction of localization due to interaction, it is also the region where the dependence on the phase $\theta$ is the largest. Unfortunately, in typical experimental situations with Bose-Einstein condensates, the phase $\theta$ is not controllable. Moreover, in the experiments the initial distribution of atoms in the lattice sites is more similar to a Gaussian than a $\delta$-function.  This suggests that, while the single-site diffusion is conceptually important and widely investigated from the theoretical viewpoint, the diffusion from a Gaussian is also interesting and worth exploring.  

\begin{figure}[b!]
\begin{center}
\includegraphics[width=0.67\columnwidth,angle=-90]{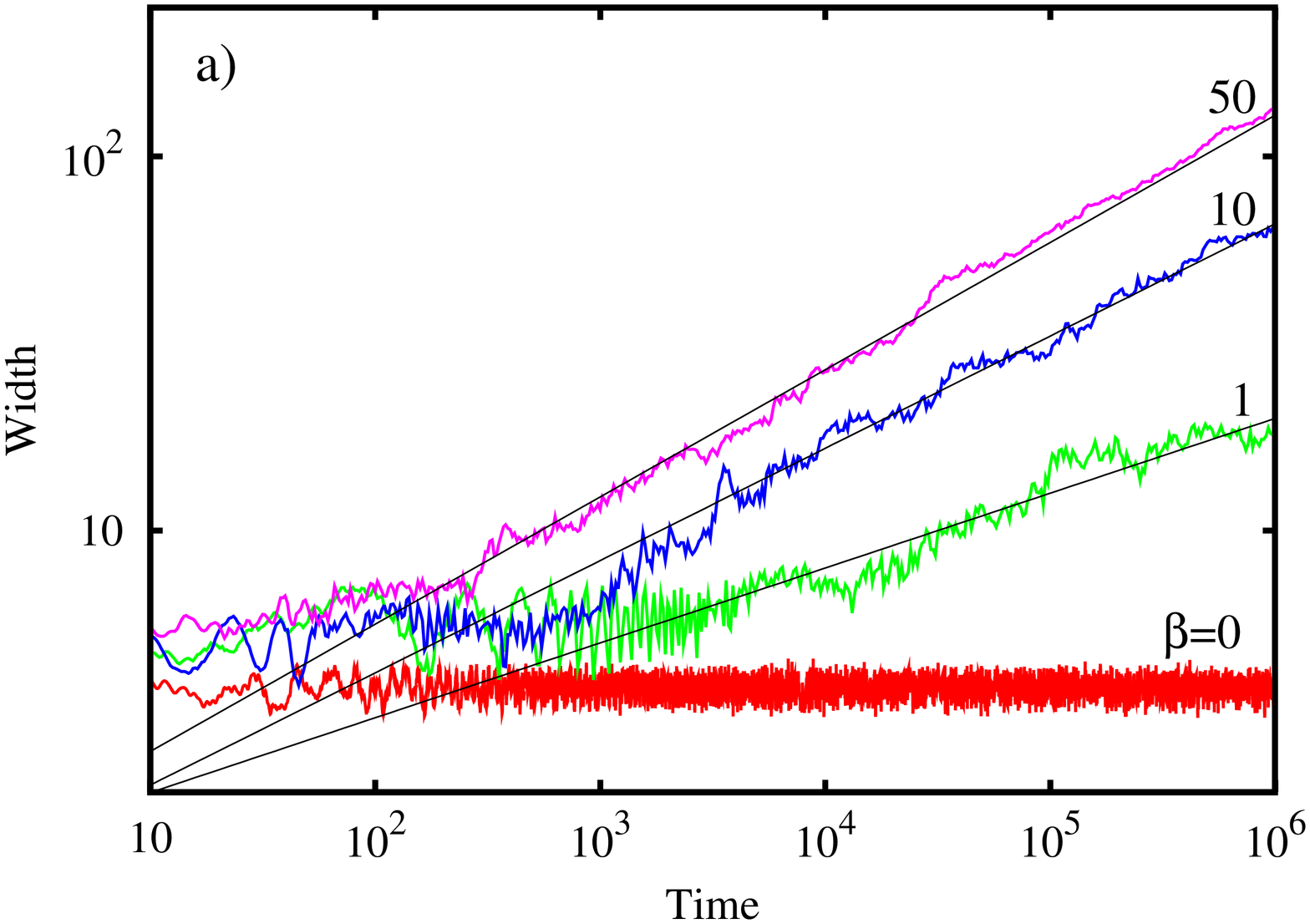}
\includegraphics[width=0.67\columnwidth,angle=-90]{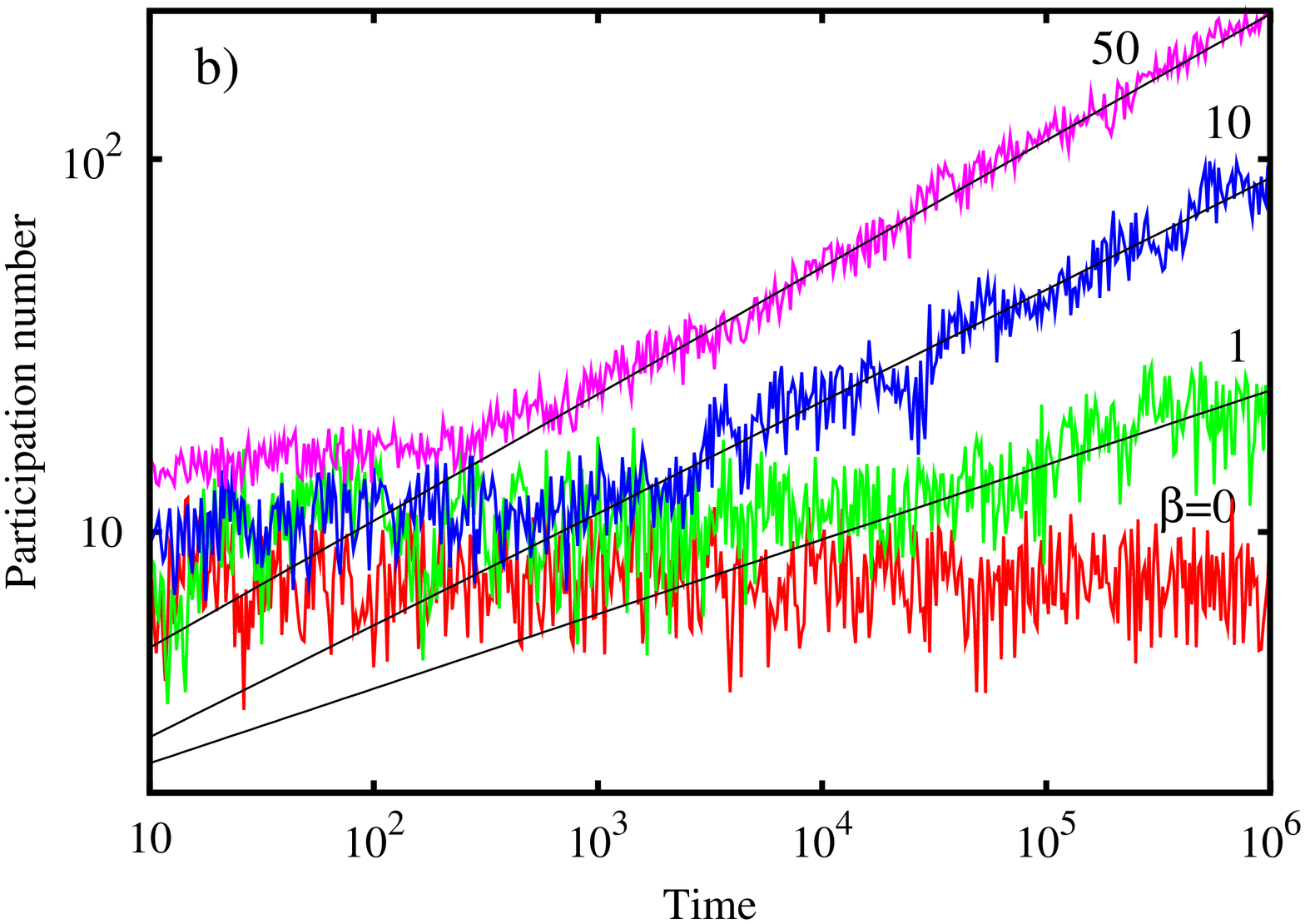}
\end{center}
\caption{(color online) Time evolution of the width of the wavepacket $w(t)$ (a) and of the participation number (b) for $\lambda=2.5$ and for an initial Gaussian wavepacket with $\sigma=5$. We compare the noninteracting case, $\beta=0$, with three interacting cases: $\beta=1, 10, 50$. The black straight lines represent a guide to the eye. Their slope is $0.2$, $0.3$ and $0.34$ and is the same in (a) and (b). These lines suggest that $\gamma_1\approx\gamma_2$ and that $\gamma$ depends on $\beta$ (see text). }
\label{fig:delocalization_beta}
\end{figure}

By repeating the same calculation of $\beta_c$ as before, but starting from a Gaussian wavepacket of width $\sigma$, we find two main results: i) if $\sigma$ is of the order of $5$ or more, the time evolution of the width, the participation number and the density distribution becomes almost independent of the phase $\theta$ for all values of $\lambda$ and $\beta$; ii) self-trapping is strongly suppressed, especially in the region $\lambda>2$. An example is shown in Fig.~\ref{fig:delocalization_beta} for $\lambda=2.5$ and an initial Gaussian of width $\sigma=5$. As one can see, self-trapping does not occur even for values of the interaction parameters of the order of $\beta\sim50$. This fact is important in view of the discussion about the delocalization induced by the interaction, which is the subject of the next section. 

\begin{figure}[t!]
\begin{center}
\includegraphics[width=0.67\columnwidth,angle=-90]{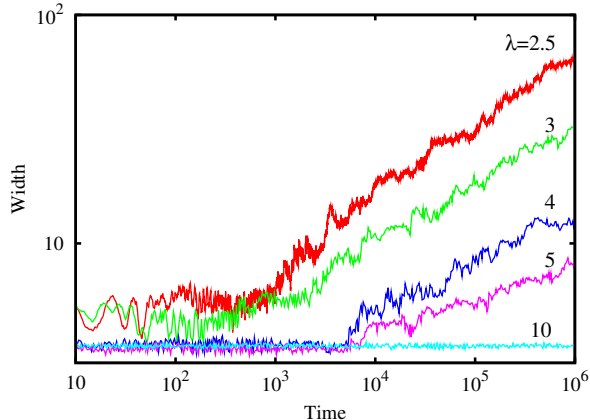}
\end{center}
\caption{(color online) Time evolution of the width of a Gaussian wavepacket with $\sigma=5$, for $\beta=10$ and different values of the disorder strength $\lambda$.}
\label{fig:delocalization_lambda}
\end{figure}
\begin{figure}[b!]
\begin{center}
\includegraphics[width=0.95\columnwidth]{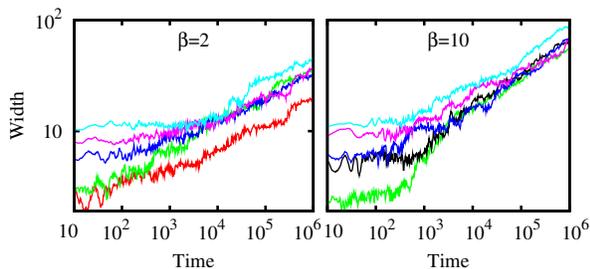}
\end{center}
\caption{(color online) Width $w(t)$ as a function of time for $\lambda=2.5$ and two different values of the interaction strength, $\beta=2$ (left column) and $\beta=10$ (right column). In the left plot we compare the time evolution for a single-site $\delta$-function (red line) and for Gaussian wavepackets with different values of the initial width, $\sigma= 2$ (green line), $6$ (blue line), $10$ (purple line), $14$ (azure line) while in the right plot we consider only Gaussian wavepackets with $\sigma=2$ (green line), $5$ (black line), $6$ (blue line), $10$ (purple line), $14$ (azure line). The curves exhibit the same asymptotic slope, which is 0.23 (left plot) and 0.3 (right plot).}
\label{fig:delocalization_sigma}
\end{figure}

\subsection{Delocalization induced by the interaction}

Let us now investigate the interplay of interaction and localization in the diffusion of a Gaussian wavepacket, in the regime where the noninteracting system is localized ($\lambda>2$) and for a choice of parameters that rules out self-trapping. In this case we find that, as the interaction is turned on, a wavepacket that was localized for $\beta=0$ starts to expand sub-diffusively. We observe an asymptotic growth of both the width $w(t)$ and the participation number $P(t)$, according to the following laws: $w(t)\sim t^{\gamma_1}$ and $P(t)\sim t^{\gamma_2}$ with $\gamma_{1,2}$ in the range $(0,0.5)$. The same effect has previously been reported for single-site diffusion in a purely random systems \cite{Pikovsky, Garcia, Flach, Skokos}. In the absence of self-trapping we find that the coefficients $\gamma_1$ and $\gamma_2$ are nearly equal, therefore in the following we will use $\gamma\approx\gamma_1\approx\gamma_2$. An example is shown in Fig.\ref{fig:delocalization_beta} where one sees the effects of the interaction on the diffusion of an initial Gaussian wavepacket with $\sigma=5$, and a disorder strength just above the localization transition, $\lambda=2.5$. The noninteracting case, which remains localized, is compared with three different values of the interaction parameter, $\beta=1$, $\beta=10$ and $\beta=50$. Already for $\beta=1$ there is an evident delocalization and this effect increases as $\beta$ is increased in the sense that $\gamma$ becomes larger and the delocalization takes place earlier. A very similar behaviour is obtained also for the diffusion from a single-site, provided the phase $\theta$ and the interaction $\beta$ are chosen in such a way to avoid self trapping (e.g., in region III of  Fig.\ref{fig:self_trapping_2} with $\theta=\pi$). 

When the disorder strength $\lambda$ is increased the localization gets more robust, in the sense that the onset of sub-diffusive spreading takes place for later times and $\gamma$ becomes smaller. For large $\lambda$ we reach a situation where the delocalization process is no longer observable within our simulation time. This is shown in Fig.\ref{fig:delocalization_lambda}, where we compare the time evolution of a Gaussian wavepacket for fixed $\beta$ and for increasing values of the disorder strength $\lambda$. Finally, in Fig.\ref{fig:delocalization_sigma} we show the behaviour of $w(t)$ as function of time for different values of the initial width of the wavepacket. In the left plot, which is for $\beta=2$, we compare a $\delta$-function wavepacket with some Gaussian cases with $\sigma= 2,\,6,\,10,\,14$  while in the right plot, which is for $\beta=10$, we consider Gaussian wavepackets with $\sigma=2,\,5,\,6,\,10,\,14$. We find that changing the width of the initial wavepacket does not affect the spreading mechanism, even in the limit $\sigma\rightarrow0$ of a $\delta$-function wavepacket.  In fact, one can see from this figure that there is no visible dependence of the asymptotic slope, $\gamma$, on the initial width.

\section{Conclusions}

The role of inter-particle interaction in the destruction of Anderson localization is rather nontrivial and is the subject of current investigations both theoretically and experimentally. Here we have considered the case of the exponential localization of a wavepacket in a 1D quasi-periodic potential. Our main goal was to fill the gap between what is predicted by the Aubry-Andr\`e model for noninteracting particles in such a potential and what can be actually observed in realistic experiments with ultracold bosons in bichromatic optical lattices. As a first step in this direction, we have studied the diffusion of a noninteracting wavepackets in a commensurate (periodic) lattice and we have compared it with the case of an incommensurate (quasi-periodic) lattice. We have shown that the spatial periodicity of the commensurate lattice plays a key role in determining the type of approach to the quasi-periodic limit in a sequence of commensurate approximants. This part of our analysis confirms that the transition from diffusion to localization observed in Ref.~\cite{Roati} can correctly be interpreted in terms of the predictions of the Aubry-Andr\`e model. A second step consists of including the atom-atom interaction. To this purpose we have numerically solved a discrete non-linear Schr\"odinger equation, which generalizes the Aubry-Andr\`e model by introducing the interaction at the mean-field level. We have simulated the dynamics of matter waves starting from either a $\delta$-function localized in a single lattice site or a Gaussian wavepacket. In the former case,  we have found that the dynamics is dominated by self-trapping processes in a wide range of parameters, even for weak interaction. Conversely, in the latter case, self-trapping is significantly suppressed and the destruction of localization by interaction is more easily observable. In particular, we find that Gaussian wavepackets, which remain localized for noninteracting particles, start to spread sub-diffusively in the presence of interaction. This result is consistent with previous predictions for interacting particles in purely random potentials. We have systematically investigated the transition from localization to diffusion as a function of the strength of both the interaction and the disorder. Our analysis is intended to stimulate further experimental work on the diffusion of atomic matter waves in bichromatic lattices.

\begin{acknowledgments}

We are indebted to C.Menotti, G.Roati and A.Smerzi for fruitful discussions. This work is partially supported by MiUR. 

\end{acknowledgments}


\begin{thebibliography}{99}
\bibitem{note-time} A linear fit of the width $w(\tau)$ at the transition time $\tau$, as a function of the spatial periodicity, gives $[w(\tau)](q_n)=aq_n+b$, with $a=0.547(4)$ and $b=0.8(2)$. The times $\tau$'s and the relative widths $w(\tau)$'s have been determined by fitting the sub-diffusive behaviour of the incommensurate case and the linear behaviours of the commensurate cases and finding the intersection points between these two fits.
\bibitem{note-expt} The time scale of experimental observation is typically less than $1000$ in our dimensionless units for the experiments of Ref.\protect\cite{Roati}, where the value $1032/862$ has been used for $\alpha$.
\bibitem{self_trapping}  When $\lambda>2$ and the interaction strength $\beta$ is of the order of, or larger than the upper bound found by energy conservation arguments we find that the self-trapping can be so strong to suppress also the growth of the width $w(t)$ (see also \cite{Skokos}).
\end{thebibliography}
\end{document}